# Investigation of the Impact of Cold Plasma Treatment on the Wettability of Medical Grade Polyvinylchloride

Edward Bormashenko[*a,] Irina Legchenkova[a], Shiri Navon-Venezia[b], Mark Frenkel[a], Yelena Bormashenko[a]

[a]Ariel University, Chemical Engineering Department, Engineering Sciences Faculty, 407000, P.O.B. 3, Ariel, Israel

[b]Department of Molecular Biology, Faculty of Natural Sciences and the Adelson School of Medicine, Ariel University, Ariel 40700, Israel

E-mail: edward@ariel.ac.il

**Abstract**

Impact of the Corona, dielectric barrier discharge and low pressure radiofrequency air plasmas on the wettability of the medical grade polyvinylchloride was investigated. Corona plasma treatment exerted the most pronounced increase in the hydrophilization of polyvinylchloride. The specific energy of adhesion of the pristine and plasma treated PVC tubing is reported. The kinetics of hydrophobic recovery following the plasma treatment was explored. The time evolution of the apparent contact angle under the hydrophobic recovery is satisfactorily described by the exponential fitting. Energy-dispersive X-ray spectroscopy of the chemical composition of the near-surface layers of the plasma treated catheters revealed their oxidation. The effect of the hydrophobic recovery is hardly correlated with oxidation of the polymer surface, which is irreversible.

1. **Introduction**

   A diversity of engineering applications calls for the modification of the surface properties of industrial polymer materials, including their adhesion and wettability. These applications include packaging, protective coatings, adhesives, textile, printing, etc. In the fields of medicine and biotechnology many products such as catheters and medical implants are completely or partly made of polymers. In contact with biological systems, compatibility of these materials is not always specified. Medical applications often demand modification of the properties of polymer surfaces. Plasma techniques supply an appropriate tool for the generation of the demanded surface properties.[1-6] Plasma modified materials fulfil the requirements for bioactivity in medicine; for example, the inclusion of antimicrobial agents (metal nano-particles, antimicrobial peptides, enzymes, etc.) in plasma modified materials (polymeric, metallic, etc.) alters them to produce superior antibacterial biomedical devices with a longer active life.[4]

Various kinds of discharges were already applied for surface modification of medical polymers including the atmospheric[7] and low pressure plasma discharges.[8] The plasma treatment creates a complex mixture of surface functionalities, which influence surface physical and chemical properties; this results in a dramatic increase in the surface energy and consequent change in the wetting behaviour of the surface.[9-12] It was demonstrated recently that the plasma treatment may result in cross-linking of polymer molecules.[13] It was reported that plasma treatment leads to the essential electrical charging of a polymer surface observed for both synthetic and biological polymers.[14-15] It is well accepted that the cold plasma treatment increases the specific surface energy of polymers, thus resulting in their pronounced hydrophilization.[16-21] At the same time the precise physico-chemical mechanism of this hydrophilization remains debatable. It was suggested that hydrophilization of organic surfaces by plasmas may be at least partially related to the re-orientation of hydrophilic moieties constituting organic molecules.[22-23] Oxidation of plasma-treated surfaces and removal of low-mass fragments present on organic surfaces also contribute to hydrophilization.[24-25]

It was reported that the specific surface energy of the cold plasma-treated polymers is decreased with time and consequently hydrophilization provided by the plasma treatment is partially lost with time.[26-30] This process is called hydrophobic recovery.[26-30] It was demonstrated experimentally that hydrophobic recovery is stipulated by the re-orientation of the polar groups constituting the polymer chains, which were oriented by the plasma treatment.[28] A model for the hydrophobic recovery due to a combination of two thermodynamically non-equilibrium processes: diffusion and molecular reorientation was suggested.[30] Our paper focuses on the estimation of impact of the cold plasma treatment on the wettability and hydrophobic recovery of the medical grade Polyvinylchloride (PVC), broadly used for manufacturing of medical catheters. Cold plasma treatment of PVC was reported by several groups,[31-34] however the experimental data related to this polymer remain sparse. We report comparative study of the impact of various plasma discharges, including the Corona discharge, dielectric barrier (DBD) discharge and radiofrequency low pressure air plasma discharge on the wettability and hydrophobic recovery of the medical grade PVC.

## 2. Experimental
### 2.1. Materials

PVC CH18 suction catheters (diameter 6 mm, length 53 cm), supplied by Unomedical ConvaTec Lim. UK, were plasma treated in our experiments. For the purposes of the goniometric measurements of wettability the catheters were flattened under $t = 80^0 C$ to and pressure of 1kPa during 40 min, followed by the cooling to ambient conditions during 1 hour.

The de-ionized water, used for the study of wettability of catheters, was purified by a synergy UV water purification system from Millipore SAS (France) and its specific resistivity was $\hat{\rho} = 18.2 M\Omega \times$ cm at 25ºC.

### 2.2. Methods

Corona plasma device (3DT, MULTIDYNE 1000, Germantown, WI, USA) consisted of a treating head that contained two hook-shaped wire electrodes. The plasma was generated under high voltage at electrode $2 \times 12$ kV and a frequency of 50 Hz at atmospheric pressure conditions, using ambient air as a carrier gas. The distance between the electrode and the flattened catheter was 2±0.1cm.

The parameters of an air low pressure radiofrequency plasma discharge were: the plasma frequency was 13.56 MHz; the power was 18 W; the pressure was 2 Torr; the times of irradiation were varied: 10, 20, 30 s; the volume of the discharge chamber was 840 cm³. Dried compressed air used for the radiofrequency plasma treatment of catheters was supplied by Oxygen & Argon Works, Ltd., Israel; moisture was less than 10 ppm, the concentration of oxygen was 20-22%.

The diffuse coplanar surface barrier discharge was created by the DBD plasma unit RPS 40, supplied by Roplass, Brno Czech Pepublic; the naximal power 40 W, frequency $\sim$ 20 KHz. The distance between the electrode and the flattened catheter was 0.5mm. The time span of the plasma treatment was varied within 15-60 s.

Apparent contact angles were established using the Ramé–Hart goniometer (Model 500). Eight measurements were taken to calculate mean apparent contact angles at ambient conditions. The apparent contact angles were taken on both sides of a droplet; the results were averaged.

Contact angle hysteresis was established with the tilted plane method. A 5 µl water resin droplet was placed on the PVC sample mounted on the glass slide. The slide was tilted until the drop began to move. The front and rear contact angles at which the droplet started to slip are regarded as the advancing $\theta_{adv}$ and receding $\theta_{rec}$ contact angles, correspondingly, as depicted in **Figure 2**. The difference $\Delta\theta = \theta_{adv} - \theta_{rec}$ is called the contact angle hysteresis.[35-39] In spite of the fact that the "tilted plane experiment" was criticized as a method for the accurate establishment of the contact angle hysteresis, it is still broadly used for its estimation.[39-40] In parallel, the contact angle hysteresis was measured with needle-syringe method. The 5 µl water droplet placed on the flattened catheter was inflated with a syringe as shown in **Figure 3A**. When the contact (triple line) was pinned and the contact angle was increased till a certain threshold value beyond which the triple line does move. This threshold contact angle $\theta_{adv}$ was regarded the advancing contact angle.[35-39] When a droplet was deflated as depicted in **Figure 3B**, its volume decreased to a certain limiting value; in parallel the contact angle decreases till a threshold value $\theta_{rec}$, interpreted as the receding contact angle.[35-39]

Chemical composition of the pristine and cold plasma treated PVC catheters was studied with SEM/EDS (scanning electron microscopy/energy dispersive spectrometry) carried out with SEM (MAIA3 TESCAN).

### 3. Results and discussion
### 3.1 Experimental study of wetting of the pristine PVC tubing

All of investigated PVC tubing (namely pristine and plasma treated by different discharges demonstrated the partial wetting regime.[35-39] This means that spreading parameter $S$,

$$S = \gamma_{SA} - (\gamma + \gamma_{SL}), \quad (1)$$

governing the wetting regime was negative.[37-39] In Eq.(1), $\gamma, \gamma_{SA}$ and $\gamma_{SL}$ are interfacial tensions at water/air, PVC/air and PVC/water interfaces, respectively. The equilibrium "as placed" equilibrium water contact angle was established at the non-treated PVC as $\theta = 93 \pm 0.5°$. This value implies the useful physico-chemical insight. Indeed, according the Young equation, defining the equilibrium (or Young) contact angle is given by:

$$cos\theta_Y = \frac{\gamma_{SA} - \gamma_{SL}}{\gamma} \quad (2)$$

where $\gamma_{SA}, \gamma_{SL}$ and $\gamma$ are the interfacial tension at the solid/air, solid/liquid and liquid/vapor interfaces correspondingly. It is recognized from the experimental data that the equilibrium water contact angle for PVC is close to $\theta \cong \theta_Y \cong \frac{\pi}{2}$. This means that $\gamma_{SA} \cong \gamma_{SL}$ takes place for PVC and this is an important conclusion, because the interfacial tension at solid/liquid interfaces is not a well-known physical value. Assuming $\gamma_{SL} \cong 35 - 39 \frac{mJ}{m^2}$,[41-42] we conclude that for PVC $\gamma_{SA} \cong 35 - 39 \frac{mJ}{m^2}$ takes place. The establishment of the equilibrium contact angle enables the calculation of the specific energy of adhesion for the water/PVC pair according to the Young-Dupre formula:

$$W \cong \gamma(1 + cos\theta_Y) \cong 70 \pm 1 \frac{mJ}{m^2} \quad (3)$$

The as-placed equilibrium contact angle does not exhaust the characterization of wettability of a solid surface. The advancing (maximal) $\theta_{adv}$ and receding (minimal) $\theta_{rec}$ contact angles should be established experimentally.[35-40] Regrettably, the values of the advancing and receding contact angles depend strongly on the experimental technique used for their establishment.[43,44] Thus, we carried the measurement in parallel with the well-known the "tilted plane" and "needle-syringe" methods, illustrated with **Figures 2-3**. The maximal and minimal contact angles emerging from both of the techniques were $\theta_{adv} = 95 \pm 1°$ and $\theta_{rec} = 74 \pm 1°$. Thus the contact angle hysteresis established for pristine PVC as high as $\Delta\theta = \theta_{adv} - \theta_{rec} \cong 21°$ was established, which is typical for the polymer surfaces.[40]

### 3.2 Influence of the plasma treatment on the wetting of PVC tubing

Plasma treatment of the PVC tubing by all kinds of the studied plasma discharges increased the hydrophilicity of the PVC catheters as shown in **Figure 4**. However, the Corona plasma treatment demonstrated the most pronounced influence on the wettability of catheters, as at it recognized from **Figure 4**. One minute plasma treatment of the tubing by the Corona plasma discharge decreased the apparent contact angle from $\theta = 93 \pm 0.5^0$ to $\theta = 30 \pm 0.5^0$, whereas the apparent contact angles attained by the radiofrequency plasma discharge and DBD discharge were $\theta \cong 50 - 60^0$ and $\theta \cong 60 - 70^0$, correspondingly. Thus, we conclude that the influence exerted by the Corona discharge on the surface properties of the PVC tubing was maximal. The increase in the energy of adhesion, following the plasma treatment as established with the Dupre formula is summarized in **Table 1**. It is noteworthy that the impact of the radiofrequency plasma discharge and DBD discharge on the wettability of PVC comes to saturation within first 15 s of the plasma treatment, whereas the influence of the Corona discharge grows slightly within the first minute of treatment, as shown in **Figure 4**.

### 3.3. Study of the hydrophobic recovery following the plasma treatment of the medical grade PVC

The change in the surface properties of polymers induced by the cold plasma treatment is lost with time.[26-30] This effect is called the "hydrophobic recovery".[26-30] Understanding of the kinetics and mechanisms of the hydrophobic recovery is crucial for applications of plasma treated polymers. The pronounced hydrophobic followed the plasma treatment of plasma treated PVC tubing as illustrated with **Figures 5-7**, representing the time evolution of the apparent contact angle measured on the plasma treated medical grade PVC for various kinds of plasma discharges. The kinetics of hydrophobic recovery is satisfactorily described by the exponential fitting, suggested in ref. 28:

$$\theta(t) = \tilde{\theta}\left(1 - e^{-\frac{t}{\tau}}\right) + \theta_0 = \theta_{sat} - \tilde{\theta}e^{-\frac{t}{\tau}}, \qquad (4)$$

where $\theta_0$ is the initial apparent contact angle, taken immediately after plasma treatment of PVC catheters, $\tau$ is the characteristic time of restoring of the contact angle, $\tilde{\theta}$ is the fitting parameter, and $\theta_{sat} = \tilde{\theta} + \theta_0$ is the saturation apparent contact angle. The numerical values of the parameters of the exponential fitting for the investigated plasma discharges are supplied in **Tables 2-4**.

A number of common features are recognized for the hydrophobic recovery of PVC tubing exposed to various plasma discharges: i) the hydrophobic recovery following the plasma treatment is well-described by the exponential fitting for all kinds of the studied discharges; ii) the saturation contact angle $\theta_{sat}$ is slightly lower than the initial apparent contact angle; in other words, the hydrophobic recovery following the plasma treatment of PVC is never complete.

On the other hand, the characteristic time of the hydrophobic recovery is largest for the radiofrequency plasma discharge, namely $\tau \cong 3.0 - 6.0$ days; it is intermediate for the Corona discharge, i.e. $\tau \cong 1.0 - 2.5$ days; and it is small for the DBD discharge, namely: $\tau \cong 0.5 - 1.5$ days. Thus, the hydrophobic recovery following the cold plasma treatment is the slowest for the radiofrequency plasma discharge, and it is rapid for the DBD plasma treatment. At this stage of our investigation we are far from the microscopic interpretation of these experimental findings, which are of a primary importance for the applications of the plasma processing of PVC, however some conclusions may be made from the energy-dispersive X-ray spectroscopy (abbreviated further EDS spectroscopy) of the plasma-treated PVC catheters, discussed in the following section.

**3.4. EDS analysis of the chemical composition of the plasma treated PVC catheters.**

EDS study of the chemical composition of the pristine and plasma-treated PVC catheters indicated the growth of the concentration of oxygen in the near surface layers of the plasma treated PVC catheters, as demonstrated in **Figure 8**. The growth of the concentration of oxygen from 7 to 9 atomic percent was registered, as shown in **Figure 8**. It should be emphasized that this tendency was kept for all kind of the investigated plasma discharges. The concentration of oxygen in the surface layer of catheters grew with the time of treatment, as shown in **Figure 8**. These results coincide with the finding reported by the other research groups[45]. It was demonstrated that when PVC is treated with an argon plasma unsaturated bonds are created at the surface.[46] It was also shown that the pendant groups of PVC are removed by the argon plasma treatment.[46] This resulted in the formation of unsaturated bonds and cross-links in the modified layer and the outermost top layer of the polymer became oxidized after exposure to air due to a reaction between long-living radicals and oxygen.[46] It is reasonable to suggest that that the similar mechanism takes place also under the cold air plasma treatment, thus, giving rise to the change in the wettability of PVC, discussed in detail in the previous sections.

It was instructive to study the surface composition of the plasma treated PVC catheters after one week of ageing under ambient conditions. It is recognized from the data presented in **Figure 9** and its comparison to the data shown in **Figure 8**, that the concentration of oxygen in the near-surface layers of the plasma-treated catheters decreased very slightly. This observation is quite expectable; indeed, oxidation of PVC emerging from the plasma treatment is an irreversible process.[45,46] Consider that the time of ageing was markedly larger that the characteristic time of the hydrophobic recovery, supplied in **Table 1**. Thus, we conclude, that the process of the hydrophobic recovery is hardly correlated with the oxidation of the polymer arising from the cold plasma treatment.

**Conclusions**

Cold plasma treatment is broadly used for modification of wettability of organic and nonorganic compounds.[1-9, 47] Influence of the plasma treatment on the surface properties of the polyvinylchloride catheters was explored with various plasma sources. We conclude that the Corona, low pressure radiofrequency and DBD plasma discharges increased essentially hydrophilicity of the PVC tubing. The Corona plasma treatment demonstrated the most pronounced influence on the eventual wettability of catheters. One-minute plasma treatment of the tubing by the Corona plasma discharge decreased the apparent contact angle from $\theta = 93 \pm 0.5^0$ to $\theta = 30 \pm 0.5^0$. The energy of water adhesion of pristine and plasma treated PVC, estimated with the Dupre formula is reported. Kinetics of the hydrophobic recovery following the plasma treatment of PVS was explored.[26-30] The hydrophobic recovery following the plasma treatment of PVC is never complete, in other words the saturation contact angle $\theta_{sat}$ is slightly lower than the initial apparent contact angle inherent for the pristine PVC. It is noteworthy that the kinetics of hydrophobic recovery following the plasma treatment is well-described by the exponential fitting suggested in ref. 28 for all kinds of the studied discharges. We established that the hydrophobic recovery following the cold plasma treatment is the slowest for the radiofrequency plasma discharge, and it is most rapid for the DBD plasma treatment of PVC. Energy-dispersive X-ray spectroscopy of the chemical composition of the near-surface layers of the catheters indicated their oxidation, emerging from the cold air plasma treatment. This oxidation was irreversible and changed very slightly under one week ageing under ambient conditions. Thus, we conclude that the phenomenon of the hydrophobic recovery is hardly correlated with oxidation of the polymer (PVC) surface,

**Acknowledgements**

This work was supported by a grant from the Israel Ministry of Science and Technology, Grant no Merkava 3-16288 supporting EB and SNV.

The authors are indebted to Mrs. N. Litvak for her kind help with the energy-dispersive X-ray spectroscopy analysis of the samples.

**Table 1**. Specific energy of adhesion, established with the Dupre formula for the pristine and plasma treated PVC tubing.

| PVC tubing | Specific Energy of Adhesion, $W, \frac{mJ}{m^2}$ |
|---|---|
| Pristine | $70 \pm 1$ |
| Corona discharge treated | $132.5 \pm 2$ |
| Radiofrequency plasma discharge treated | $111.5 \pm 5$ |
| DBD plasma discharge treated | $101.5 \pm 5$ |

**Table 2**. Parameters of the exponential fitting of the kinetics of hydrophobic recovery following the Corona plasma treatment of PVC.

| Exposure time, s | $\theta_{Sat}$, ° | $\tilde{\theta}$, ° | $\tau$, days |
|---|---|---|---|
| 15 | 73±2.8 | 45±3.7 | 2.2±0.5 |
| 30 | 60±1.6 | 25±2.1 | 2.2±0.5 |
| 45 | 60±1.2 | 27±2.6 | 1.2±0.3 |
| 60 | 62±5.3 | 34±5.7 | 2.7±1.3 |

**Table 3**. Parameters of the exponential fitting of the kinetics of hydrophobic recovery following the radiofrequency plasma treatment of PVC.

| Exposure time, s | $\theta_{Sat}$, ° | $\tilde{\theta}$, ° | $\tau$, days |
|---|---|---|---|
| 15 | 87±5.0 | 36±4.8 | 3.1±1.2 |
| 30 | 97±6.8 | 47±6.2 | 6±1.6 |
| 45 | 88±8.7 | 45±7.7 | 4.3±1.9 |
| 60 | 78±5.7 | 62±6.2 | 2.7±0.8 |

**Table 4**. Parameters of the exponential fitting of the kinetics of hydrophobic recovery following the DBD treatment of PVC.

| Exposure time, s | $\theta_{Sat}$ | $\tilde{\theta}$ | $\tau$ |
|---|---|---|---|
| 15 | 86±1.2 | 27±2.7 | 0.7±0.2 |
| 30 | 88±0.2 | 13±0.3 | 1.3±0.1 |
| 45 | 83±2.2 | 31±5.6 | 0.5±0.3 |
| 60 | 85±2.5 | 17±3.4 | 1.5±0.1 |

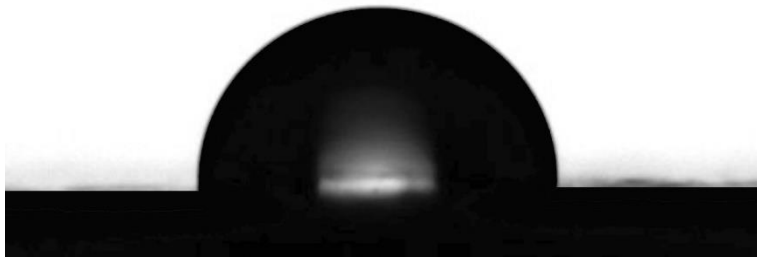

**Figure 1**. Wettability of the pristine PVC catheter is illustrated. The apparent contact angle is close to $\frac{\pi}{2}$.

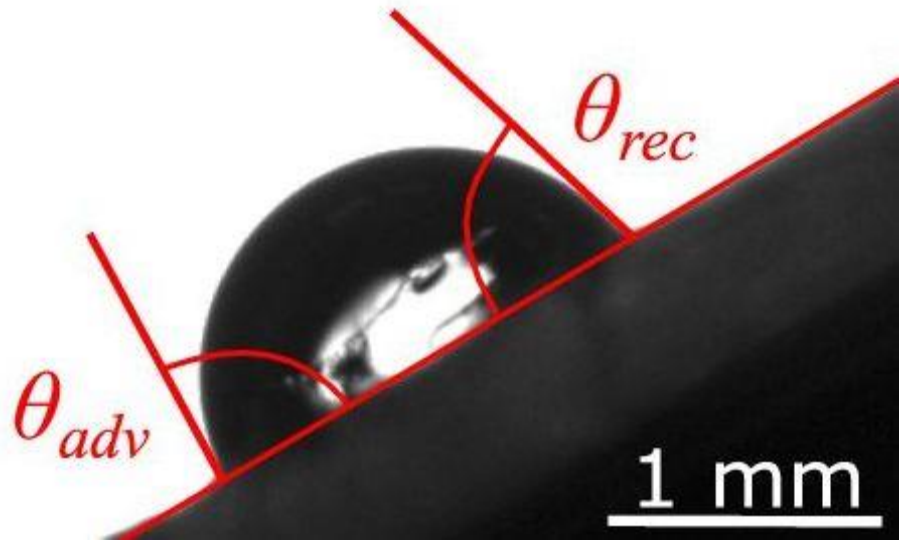

**Figure 2**. Water droplet placed on the tilted PVC surface is shown. The advancing $\theta_{adv}$ and receding $\theta_{rec}$ contacting angles are depicted.

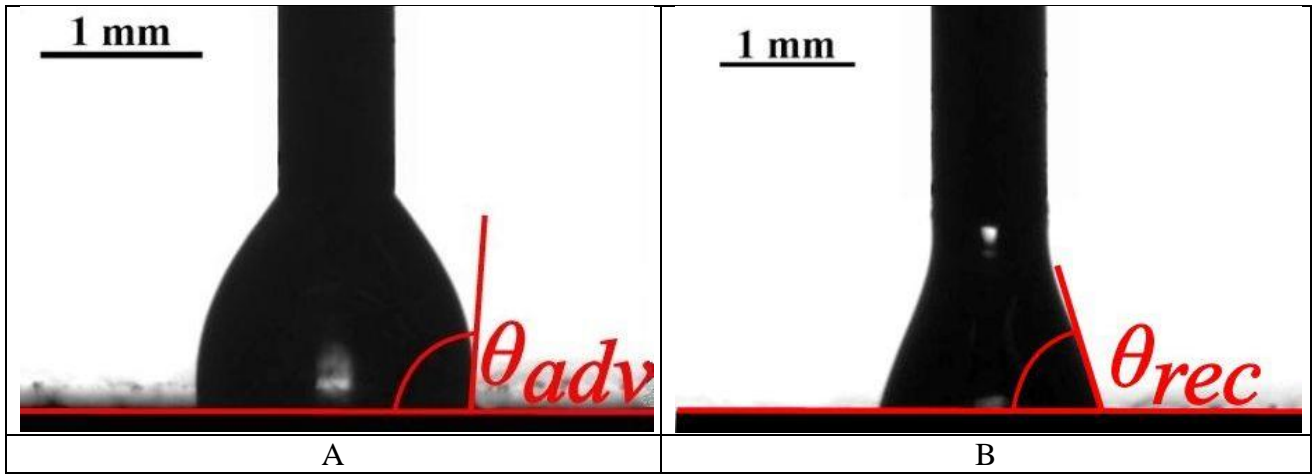

**Figure 3**. Establishment of the contact angle hysteresis on the pristine PVC catheter is demonstrated.
A. Measurement of the advancing contact angle is shown. B. Measurement of the receding contact angle is depicted.

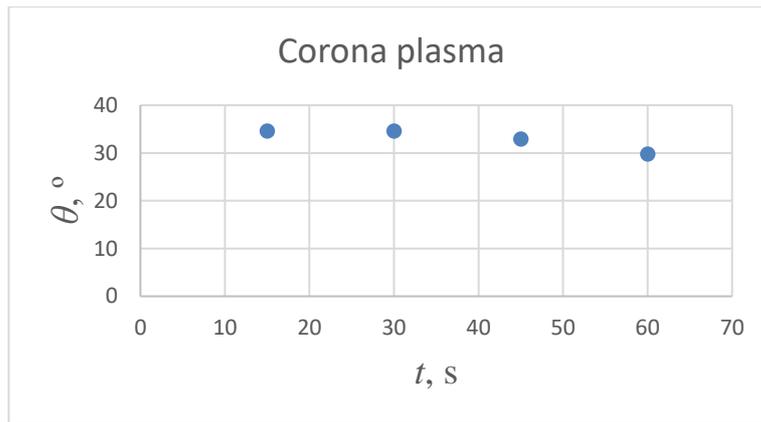

A

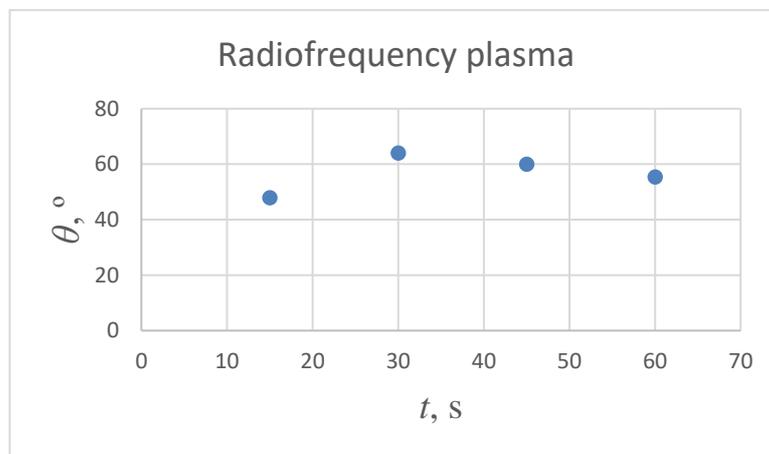

B

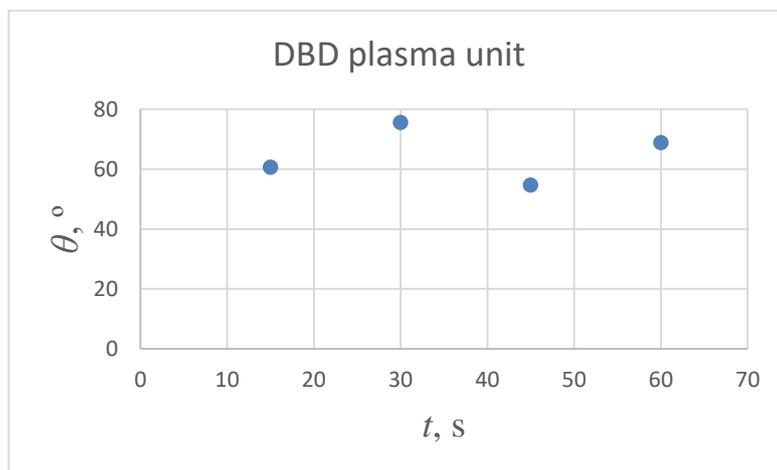

C

**Figure 4**. Impact of the dependence of the equilibrium apparent contact angle on time of plasma treatment is presented. A. Treatment with the Corona plasma discharge. B. Treatment with the Radiofrequency plasma device. C. Treatment with the DBD plasma unit.

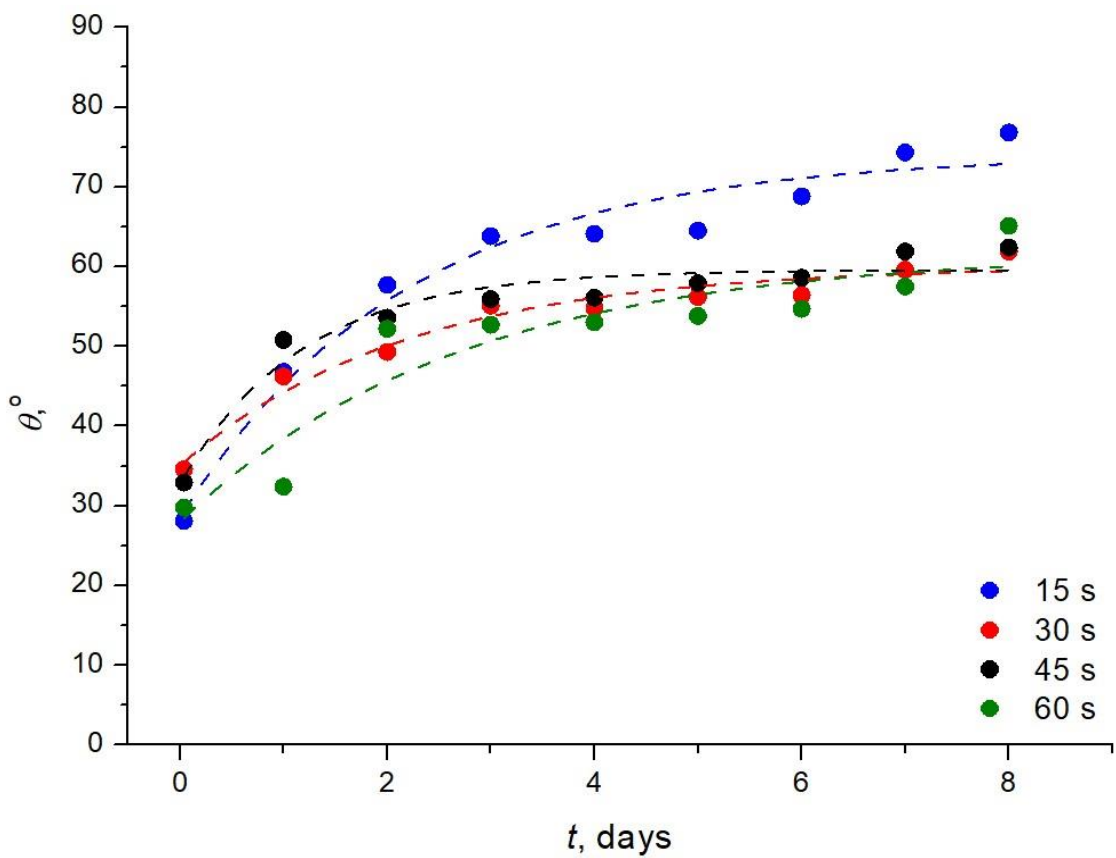

**Figure 5**. Kinetics of the hydrophobic recovery of PVC following Corona plasma treatment is shown. Blue circles correspond to the 15s treatment; red circles correspond to the 30 s treatment; black circles correspond to 45 s treatment; green circles correspond to 60 s treatment. The dashed line corresponds to the exponential fitting with Eq. 4. The parameters of fitting are supplied in **Table 2**.

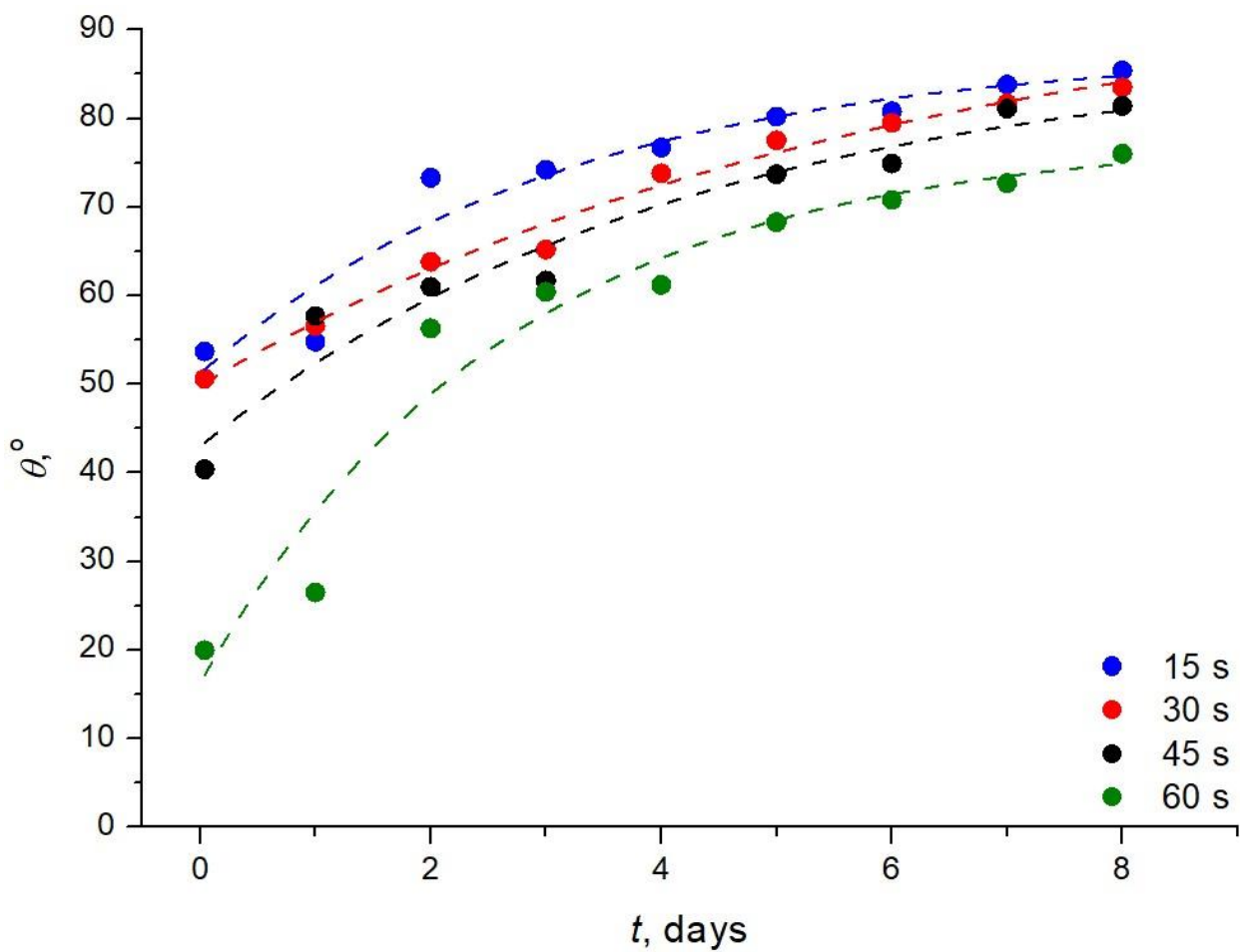

**Figure 6**. Kinetics of the hydrophobic recovery of PVC following the radiofrequency plasma treatment is depicted. Blue circles correspond to the 15s treatment; red circles correspond to the 30 s treatment; black circles correspond to 45 s treatment; green circles correspond to 60 s treatment. The dashed line corresponds to the exponential fitting with Eq. 4. The parameters of fitting are supplied in **Table 3**.

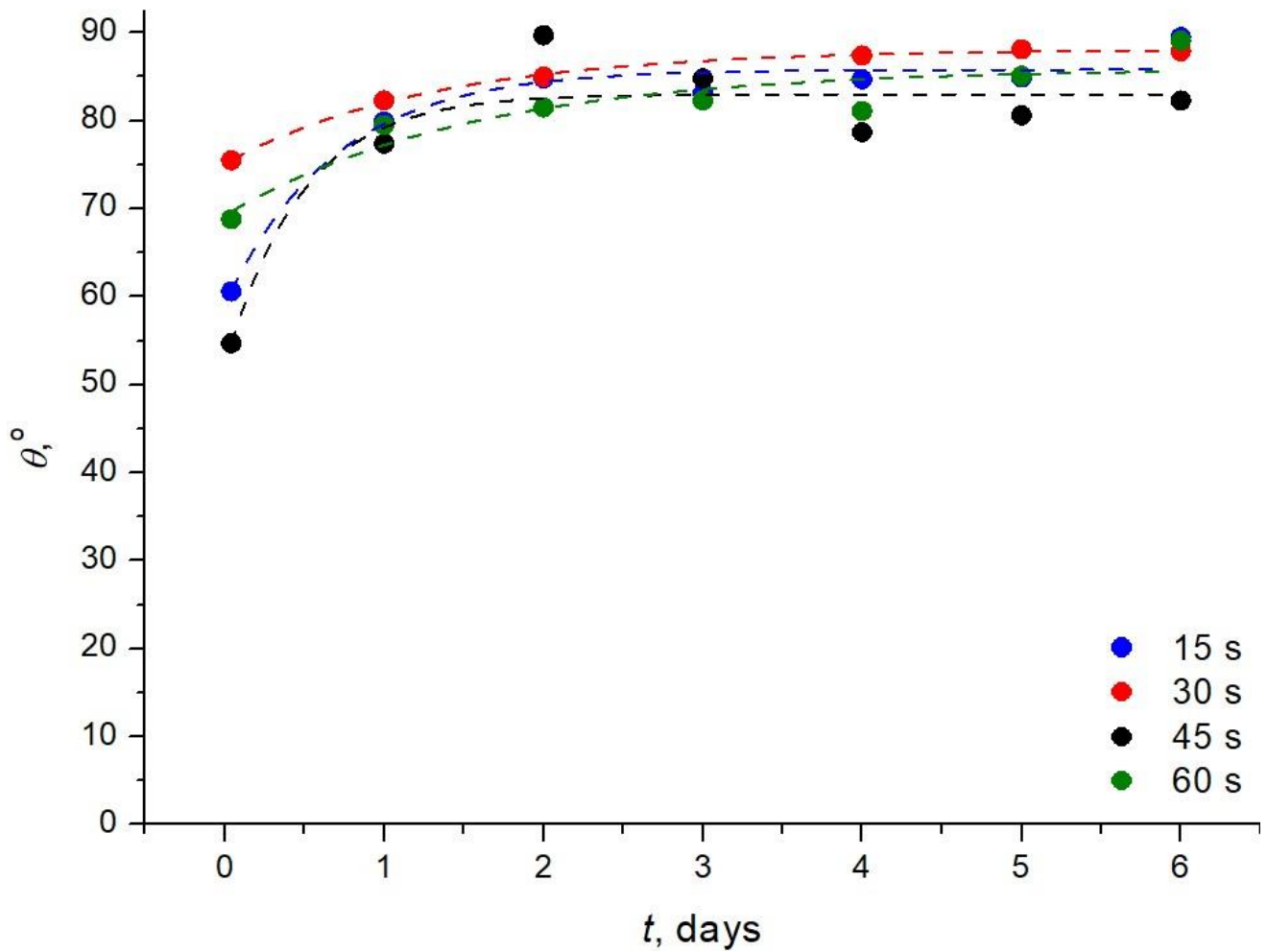

**Figure 7**. Kinetics of the hydrophobic recovery of PVC following the DBD plasma treatment is depicted. Blue circles correspond to the 15s treatment; red circles correspond to the 30 s treatment; black circles correspond to 45 s treatment; green circles correspond to 60 s treatment. The dashed line corresponds to the exponential fitting with Eq. 4. The parameters of fitting are supplied in **Table 4**.

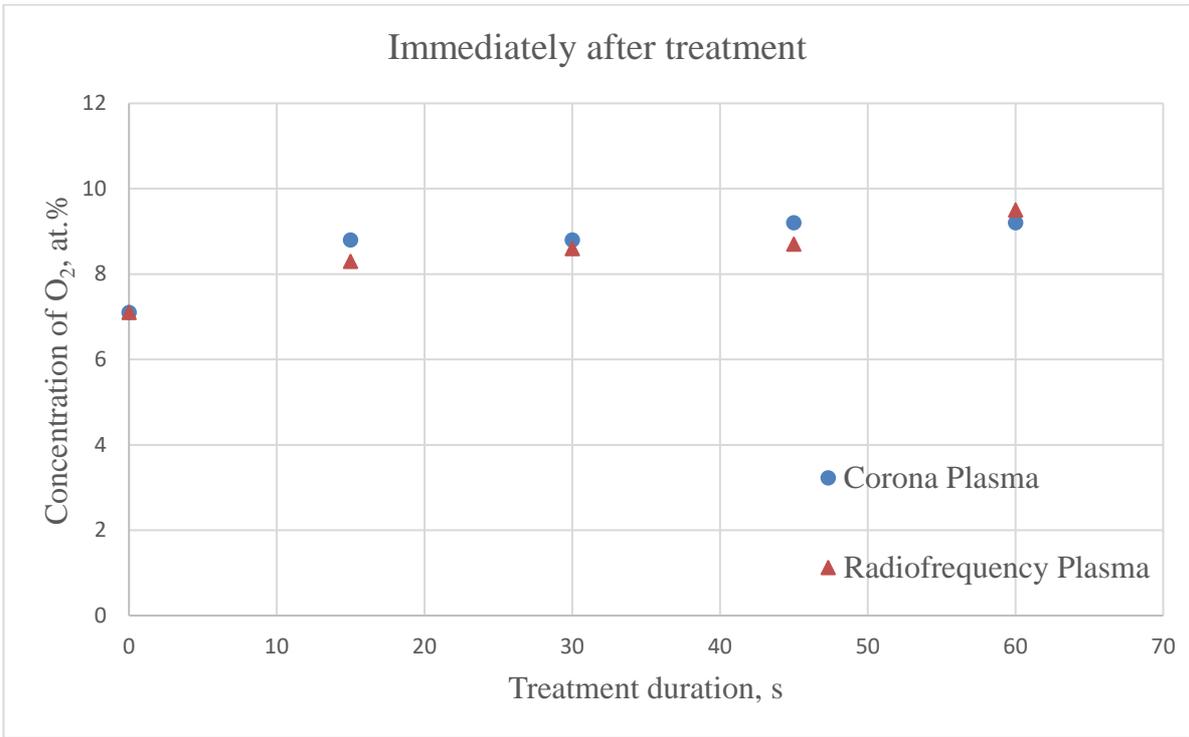

**Figure 8**. The surface concentration of oxygen (atomic percent) in the surface layer of catheters *vs*. the duration of plasma treatment as established immediately after the treatment with SEM/EDS spectroscopy is depicted.

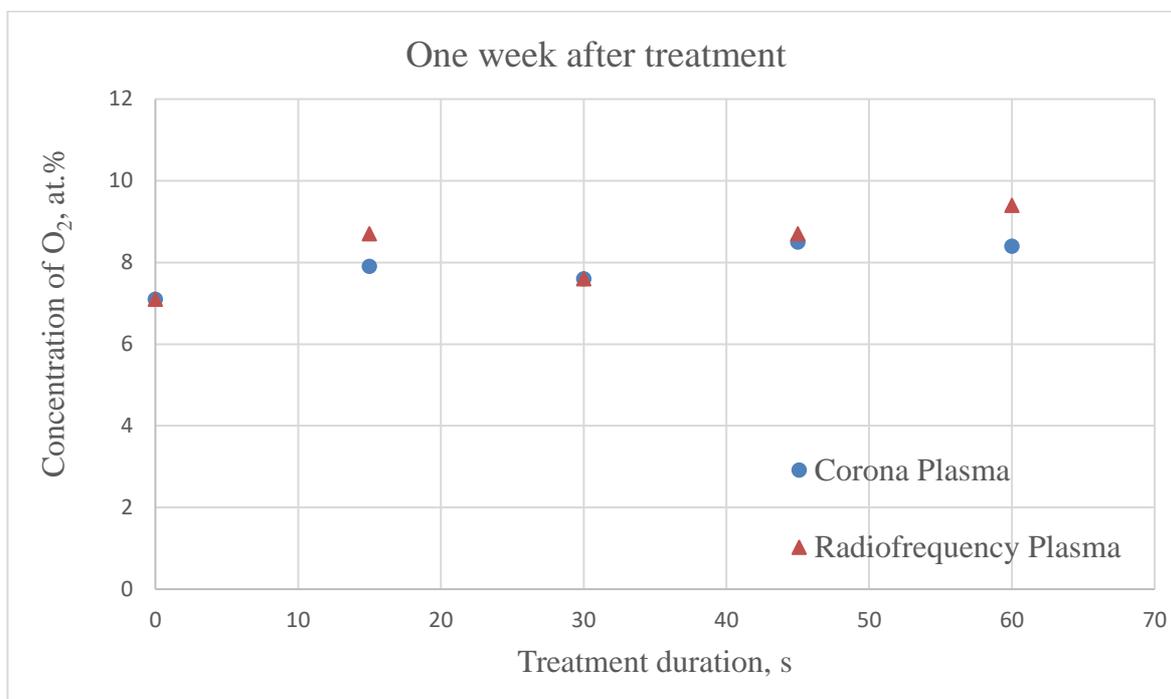

**Figure 9**. The surface concentration of oxygen (atomic percent) in the surface layer of catheters *vs*. the time of plasma treatment as established with SEM/EDS spectroscopy one week after the treatment is presented.